\documentclass[aps,prb,superscriptaddress,amsmath,amssymb,floatfix,twocolumn,showpacs]{revtex4-1}
\usepackage{graphicx}
\usepackage{amsbsy}
\usepackage{txfonts}
\usepackage{xcolor}
\usepackage{times}
\usepackage{subfigure}
\usepackage{bm}
\usepackage{mathrsfs}
\usepackage{hyperref}
\usepackage{natbib}
\usepackage{bbold}
\usepackage{empheq}
\DeclareMathOperator{\Tr}{Tr}
\usepackage{filecontents}
\usepackage{float}

\newcommand{\QQ}{\mathbf{Q}}
\newcommand{\MM}{\mathbf{M}}
\newcommand{\rr}{\mathbf{r}}
\newcommand{\ud}{\mathrm{d}}

\allowdisplaybreaks[1]

\begin{document}

\title{Orbital Nematic Order and Interplay with Magnetism in the Two-Orbital Hubbard Model}

\author{Zhentao Wang}
\author{Andriy H. Nevidomskyy}
\affiliation{Department of Physics and Astronomy, Rice University, Houston,
TX 77005}
%\pacs{75.25.Dk, %(Orbital, charge, and other orders, including coupling of these orders)
%74.70.Xa, %(Pnictides and chalcogenides)
%74.25.-q %(Properties of superconductors)
%}

\begin{abstract}
Motivated by the recent angle-resolved photoemission spectroscopy (ARPES) on FeSe 
and iron pnictide families of iron-based superconductors, we have studied the 
orbital nematic order and its interplay with antiferromagnetism within the two-orbital Hubbard model.
We used random phase approximation (RPA) to calculate the dependence of the orbital and magnetic susceptibilities on the strength of 
interactions and electron density (doping). To account for strong electron correlations not captured by RPA, we further employed non-perturbative variational cluster  approximation (VCA) capable of capturing symmetry broken magnetic and orbitally ordered phases.
Both approaches show that the electron and hole doping affect the two orders differently. 
While hole doping tends to suppress both magnetism and orbital ordering, 
the electron doping suppresses magnetism faster. Crucially, we find a realistic parameter regime  for moderate electron doping that stabilizes orbital nematicity in the absence of long-range antiferromagnetic order. This is reminiscent of the non-magnetic orbital nematic phase observed recently in FeSe and a number of iron pnictide materials and raises the possibility that at least in some cases, the observed electronic nematicity 
may be primarily due to orbital rather than magnetic fluctuations.
\end{abstract}

\maketitle

\section{Introduction}\label{Sec:Intro}
Nematicity, defined as spontaneous breaking of the four-fold rotational $C_4$ symmetry down 
to $C_2$, has been recently observed in the electronic properties of
Fe-based superconductors\cite{Kamihara08,Ren08}. Experimental efforts, including 
transport measurements\cite{Chu10,Tanatar10,Blomberg11,Kuo11,Ying11,Chu12,Jiang13,Ishida13}, 
optical conductivity\cite{Tanatar10,Nakajima11,Dusza12,Nakajima12}, 
scanning tunneling microscopy\cite{Chuang10,Zhou11,Allan13}, 
neutron scattering\cite{Harriger11,Luo13, Lu14},
quantum oscillations\cite{Terashima11}, magnetic torque measurements\cite{Kasahara12}, 
and angle-resolved photoemission spectroscopy (ARPES) measurements\cite{Shimojima10,Yi11,Shimojima14}, 
have reported the electronic in-plane anisotropy, 
mostly in the compounds of the BaFe$_2$As$_2$ (122) family, even at temperatures higher than 
the lattice structural transition\cite{Chu10, Chu12, Kasahara12, Lu14, Yi11, Shimojima14}. Upon electron doping, the 
resistivity anisotropy is first enhanced in the underdoped region, but then suppressed 
upon further doping into the superconducting region\cite{Chu10}; while hole doping 
appears to suppress the resistivity anisotropy even while below the structural phase transition\cite{Ying11}. 

Origins of the observed anisotropy have been discussed in the context of 
lattice, magnetic and orbital fluctuations\cite{Fang08-J1J2K,Xu08-J1J2,Dai09,Lee09-LDA+U,Lv09,Chen09,Kruger09,
Chen10-5orbital-mf,Lv10,Valenzuela10,Kontani11}, and 
there are also debates whether the nematicity is intrinsic or if it comes from the anisotropic 
impurity scattering\cite{Ishida13,Allan13,Kuo13}. Recent resistivity $\rho$ measurements under fixed strain\cite{Chu12} $\delta=(a-b)/(a+b)$ ($a$ and $b$ are the lattice constants) have detected divergent nematic susceptibility $\ud \rho/\ud\delta$, proving that the nematicity is of electronic origin rather than due to an elastic lattice instability. 
%In the contents of interplay between the CAF and orbital polarization, Daghofer et al~\cite{Daghofer10-3orbital-mf} concluded from their theoretical analysis that the orbital polarization of the Fermi surface measured in ARPES\cite{Shimojima10} stems primarily from the antiferromagnetic order, and not the ferro-orbital order (the latter is small, a few percent in magnitude).

One possible mechanism behind this electronic nematicity is the so-called 
Ising-nematic (also referred to as ``spin nematic'') scenario, based on the 
discrete Ising symmetry breaking between two columnar antiferromagnetic 
(CAF) ordering wave-vectors, $\QQ_1 = (\pi,0)$ and $\QQ_2 = (0,\pi)$. 
In the ordered CAF phase, the magnetization $\MM(r) = \MM_1 e^{i\QQ_1\cdot \rr} 
+ \MM_2 e^{i\QQ_2\cdot\rr}$ breaks the $C_4$ symmetry whenever $M_1 \neq M_2$, 
and neutron scattering finds either $M_1 = 0$ or $M_2 = 0$ in the iron 
pnictides\cite{Cruz08, Harriger11}. The nematic order parameter is 
defined as the difference in spin correlations along the $\hat{x}$ and 
$\hat{y}$ axes: 
\begin{equation}
\psi = \sum_i|\MM(i)\cdot \MM(i+\hat{x}) - \MM(i)\cdot \MM(i+\hat{y})|\propto |M_1|^2 - |M_2|^2. \label{Eq.psi}
\end{equation}
However, the $C_4$ symmetry can be spontaneously broken even above the 
N\'eel temperature ($T_N$) due to anisotropic spin fluctuations, as 
was first pointed out by Chandra, Coleman and Larkin\cite{CCL}, and 
later applied to the iron pnictides\cite{Fang08-J1J2K,Xu08-J1J2,Dai09, 
Fernandes11, Fernandes12-sst}. Recently, these anisotropic spin 
fluctuations have been imaged directly\cite{Lu14} using the inelastic 
neutron scattering in uniaxial-pressure detwinned samples of BaFe$_{2-x}$Ni$_x$As$_2$.

On the other hand, it has been proposed that the electronic nematicity 
may stem from unequal population of the $d_{xz}$ and $d_{yz}$ orbitals, 
resulting in the ferro-orbital ordering~\cite{Kruger09, Lee09-LDA+U, 
Lv09, Chen09, Chen10-5orbital-mf, Lv10}. The order parameter is the 
orbital polarization $p=\langle n_{xz}-n_{yz}\rangle$, which explicitly 
breaks the $C_4$ symmetry. Experimentally, polarization-dependent 
ARPES found 
orbitally-polarized Fermi surfaces inside the CAF phase of the parent 
compound BaFe$_2$As$_2$\cite{Shimojima10}.  The unambiguous 
 splitting of Fe $d_{xz}$ and $d_{yz}$ orbitals has also been observed 
 by ARPES above the structural transition temperature $T_s$ in the 
 detwinned samples of lightly Co-doped\cite{Yi11} and 
 P-doped\cite{Shimojima14} BaFe$_2$As$_2$.  
 
The driving force for the electronic nematic transition is very difficult to determine because of the strong coupling between internal spin and orbital degrees of freedom. Indeed, on the symmetry grounds, Landau free energy will contain a linear coupling between the Ising-nematic order parameter $\psi$ and the orbital ordering $p$:
\begin{equation}
\Delta F = \eta\, p\cdot \psi \;\propto\; \eta\, p\cdot \left\langle M_1^2 - M_2^2 \right\rangle.\label{Eq.free}
\end{equation}
 Independent of the sign of the coupling constant $\eta$, such a term in the free energy will result in a non-vanishing value of the orbital polarization whenever $\psi$ takes on a non-zero value, and the other way round. 
It has been proposed that this ``chicken and egg'' problem may in principle be resolved by comparing the rates of divergence in the orbital and Ising-spin susceptibilities on approaching the transition,\cite{Fernandes14} however this approach would only work provided the coupling constant $|\eta|$ is not too large. 
In this study, we aim to address a different question, namely, is it possible, and under what conditions, to stabilize a static ferro-orbital order in the absence of long-range magnetic ordering? 

In order to disentangle the effects of the magnetic and orbital ordering, it is useful to consider the compounds where the two phases are clearly separated. One such example is NaFeAs of the 111 family with $T_s=53$~K significantly higher than the N\'eel temperature $T_N=40$~K. 
 Recent ARPES measurements in NaFeAs  indicate\cite{Yi12,Zhang12} that the orbital order develops exactly at or slightly above $T_s$ and appears to trigger the antiferromagnetic order at the lower temperature $T_N$, due to the nesting of the two-fold anisotropic Fermi surface~\cite{Yi12}. One is tempted therefore to interpret this result based on the orbital-driven scenario\cite{Yi12}. Intriguingly, scanning tunneling spectroscopy finds evidence of local electronic nematicity up to temperatures twice $T_s$, in the nominally tetragonal phase\cite{Rosenthal14}.
% and in the Co-doped NaFeAs {Wang12,Wright12,Rosenthal14}. %(NaFeAs)

An even clearer signature of the orbital ordering in the absence of 
magnetism is observed in stoichiometric FeSe, which undergoes a 
structural transition at around $T_s\approx90$~K without any sign of 
the antiferromagnetic ordering\cite{McQueen09, Bendele10}. Recent 
ARPES measurements\cite{Shimojima14-FeSe} on FeSe show a clear 
splitting of $\sim50$ meV between the energies of the Fe $d_{xz}$ 
and $d_{yz}$ orbital bands, which sets in at about $T_s$. Importantly, 
this measured energy splitting is more than five times larger than 
expected from density-functional theory (DFT) calculations considering 
the orthorhombic lattice distortion alone\cite{Shimojima14-FeSe}. The likely orbital nature of the structural transition in FeSe is also corroborated by recent nuclear magnetic resonance (NMR)~\cite{Baek14, Boehmer14} and shear modulus measurements\cite{Boehmer14}.

%It has been thus proposed that the structural transition in FeSe is  driven by the orbital ordering, rather than by the Ising-nematic scenario\cite{Shimojima14-FeSe}.

The above experimental observations raise a possibility that it may 
indeed be possible to stabilize ferro-orbital order in the absence of 
long-range magnetic ordering.
It is the purpose of this article  to address 
this question theoretically within the minimal two-orbital 
model.\cite{Raghu08,Daghofer08} 
While the two-orbital model is known to have a number of limitations in describing the iron-based superconductors (for instance, resulting in a wrong number of Fermi pockets and missing the $d_{xy}$ orbital contents on the Fermi surface), the primary reason for using this model here is its conceptual simplicity. We do not presume that such a simple model can describe the realistic electronic properties of, e.g. FeSe or LiFeAs. Rather, the question we aim to address can be phrased as follows -- what is the minimal theoretical model (whether or not applicable to the iron pnictides) that can support orbital nematic ordering in the absence of magnetism? In this article, we show that the two-orbital model is sufficient to describe this physics under realistic values of intra-orbital and inter-orbital Coulomb repulsion. Generalization of these results to a realistic five-orbital model of the iron pnictides or iron chalcogenided will be the subject of future work.

We used the combination of 
RPA and non-perturbative quantum cluster calculations (VCA) to study the effect of electron interactions and doping on both the antiferromagnetism and ferro-orbital ordering.  
Within the two-orbital model, we find that the orbital nematic order strongly depends on 
inter-orbital Hubbard repulsion and Hund's coupling 
term, and its dependence on electron and hole doping is not symmetric. 
In the undoped and hole-doped system, we find that orbital order 
coexists with magnetic order, as observed 
%\red{(Actually not exactly, in the paper by \onlinecite{Ying11} mentioned in the first paragraph, hole doping 
%immediately suppresses nematiticy, I don't know if there are newer experiments  opposing that)} 
%% A.N.: I've commented on the paper by Ying (2011) in the Discussion section
ubiquitously in the 122 family of iron pnictides.
Our key finding is that sufficient electron doping stabilizes the orbital 
nematic phase while suppressing the antiferromagnetic ordering, suggestive of the 
experimental observations of orbital order in Co-doped LiFeAs\cite{Miao14} and FeSe\cite{Shimojima14-FeSe,Baek14,Bohmer14}.

This paper is organized as follows: In Section~\ref{Sec:Model} we introduce
the two-orbital Hubbard model as a starting point for our calculations; in Section~\ref{Sec:RPA}
 and \ref{Sec:VCA} we study the orbital nematic order and 
its interplay with the magnetic order by using RPA and VCA methods, 
respectively; in Section~\ref{Sec:Con}, we discuss our results in 
the context of the related theoretical and experimental work, 
and we finally draw the conclusions in Section~\ref{Sec:Con2}.

\section{Model}\label{Sec:Model}
To study the orbital nematic order in Fe-based superconductors, we start from 
the minimal two-orbital Hubbard model capturing the itinerant electrons with onsite 
electron-electron interactions:
\begin{equation}
\begin{split}
H&=H_{0}+U\sum_{i,\alpha}n_{i\alpha\uparrow}n_{i\alpha\downarrow}+
  (U^{\prime}-\frac{J}{2})\sum_{i,\alpha<\beta}n_{i\alpha}n_{i\beta} \\
  &\quad -2J\sum_{i,\alpha<\beta}\bm{S}_{i\alpha}\cdot\bm{S}_{i\beta}
  +J^{\prime} \sum_{i,\alpha<\beta} (c_{i\alpha \uparrow}^\dagger 
  c_{i\alpha \downarrow}^\dagger c_{i\beta \downarrow} c_{i\beta \uparrow}+h.c.)
\end{split}
\label{eq.2orbital}
\end{equation}
Here $i$ is the site label, $\alpha,\beta \in \{xz,yz\}$ stand for the orbital indices. 
$U$ and $U^\prime$ are respectively the intra- and inter-orbital Hubbard 
interaction, $J$ stands for the Hund's coupling and $J^\prime=J$ 
stands for the pair hopping term. 
$H_0$ is the non-interacting two-orbital Hamiltonian from Ref.~\onlinecite{Raghu08}:
\begin{equation}
H_0 = \sum_{\bm{k} \sigma} \psi_{\bm{k} \sigma}^{\dagger} [\epsilon_+ (\bm{k}) \mathbb{1} + \epsilon_- (\bm{k}) \tau_3
      +\epsilon_{xy} (\bm{k}) \tau_1] \psi_{\bm{k} \sigma}
\label{eq:H0}
\end{equation}
with $\psi_{\bm{k} \sigma}^{\dagger}=[ c_{xz,\sigma}^{\dagger} (\bm{k}), 
c_{yz,\sigma}^{\dagger} (\bm{k}) ]$, where $\sigma$ is the spin label, and
\begin{equation}
\begin{split}
\epsilon_\pm (\bm{k}) &= \frac{\epsilon_x (\bm{k}) \pm \epsilon_y (\bm{k})}{2} \\
\epsilon_x (\bm{k}) &= -2 t_1 \cos k_x -2 t_2 \cos k_y -4 t_3 \cos k_x \cos k_y \\
\epsilon_y (\bm{k}) &= -2 t_2 \cos k_x -2 t_1 \cos k_y -4 t_3 \cos k_x \cos k_y \\
\epsilon_{xy} (\bm{k}) &= -4 t_4 \sin k_x \sin k_y
\end{split}
\label{eq:eps_k}
\end{equation}
The values of the hopping parameters have been kept fixed throughout the 
context: $t_1=-0.25eV, t_2=0.325eV, t_3=t_4=-0.2125eV$. 
%Chemical potential 
%$\mu$ was adjusted self-consistently within VCA calculation 
%to keep the system at fixed filling, and within the RPA 
%$\mu=0.3625eV$ was fixed to ensure half-filling, assuming that
%the chemical potential does not change much as a function of the interaction 
%strength $U$ and $U^{\prime}$ in the weak-coupling regime.

  The reason behind studying the two-orbital (as opposed to the full five-orbital) 
  model is  the universally accepted fact that the major contribution to 
  the Fermi surface comes from the iron $t_{2g}$ orbitals ($d_{xz}, d_{yz}, 
  d_{xy}$), whereas the $e_g$ orbital weight is very small\cite{Graser09, 
  Daghofer10}. Additionally, the $d_{xz}$ and $d_{yz}$ orbitals carry most 
  of the spectral weight\cite{Graser09} and the $d_{xy}$ orbital can thus 
  be neglected in the first approximation. While it is true that in order 
  to obtain \emph{all} the Fermi pockets observed in ARPES, one needs 
  to consider all 5 Fe orbitals\cite{Kuroki08}, here we  chose to focus 
  on the two-orbital model in the hope  that it captures  the salient 
  features of nematicity in the iron pnictides,  especially because the
    ferro-orbital nematic order only affects the two $d_{xz}$ and $d_{yz}$ orbitals 
   in question. 
   We expect  that our central results will remain unaltered 
   upon inclusion of the other orbitals, however demonstrating this explicitly will be the subject of future work. We note that the present approach is similar in spirit to the 
   weak-coupling approaches~\cite{Fernandes10-C66, Fernandes12-sst, Fernandes12-prl, Chubukov12} 
   which start from the band picture of a hole pocket around the $\Gamma$-point 
   and electron pockets in the corners of the Brillouin zone, in a sense 
   that these approaches also deal with a reduced Hilbert space of typically 
   two bands (irrespective of their orbital contents). Nevertheless, even such a simplified  two-band approach is known to produce reliable results 
   which are largely unaffected by inclusion of other bands~\cite{Chubukov12}.
   
 We point out that another, more pragmatic reason for limiting the present consideration to two orbitals is because the five-orbital model is well beyond the computational demands of the state-of-the-art variational cluster approximation used in this work (see section \ref{Sec:VCA} for more detail). Studying antiferromagnetism necessitates the use of a 4-site cluster, which when combined with 5 orbitals per Fe site, would result in an effective 20-``site'' Hubbard model that lies well beyond the present limits of either the exact diagonalization or continuous-time quantum Monte Carlo solver~\cite{Gull11} used in VCA or in cluster-DMFT.

The inter- and intra-orbital interaction strengths in Eq.~(\ref{eq.2orbital}) 
are not independent of each other. In the atomic limit, a well known relation 
$U^\prime=U-2J$ holds, which ensures orbital rotational invariance\cite{Dagotto01}. 
In a solid, the electron-electron interactions are screened, meaning that 
the above relation between $U^\prime$ and $U$ may not be obeyed exactly. 
Below, we shall investigate the phase diagram of the model Eq.~(\ref{eq.2orbital}) 
treating $U^\prime$ and $U$ as independent parameters. However later on, 
when studying the effect of interactions on nematicity and antiferromagnetism, 
we shall use the relation  $U^\prime=U-2J$ which we expect to hold approximately in the iron pnictides.
 The values of interactions for the 122 Fe-pnictides in the two-orbital 
model are approximately $U=2$~eV, $U^\prime=0.6$~eV, $J=0.7$~eV. Note that 
we chose the interaction $U$ to be somewhat lower than $U=2.7$~eV calculated within the 
\emph{ab initio} constrained-RPA scheme~\cite{Aichhorn09},
% we should not cite \cite{Kutepov10}, b/c they used GW and determined U to be very large (U~F0, where the so-called Slater integral F0 is between 7 and 8 eV). I think those values of F0 are pertinent to the DMFT+GW method, but not to what we do here.
which is consistent with the smaller effective bandwidth when considering 
only $d_{xz}$ and $d_{yz}$ orbitals in Eq.~(\ref{eq.2orbital}), compared to the width of all 5 Fe bands. When we 
attempted to use larger values of $U\gtrsim 2.5$~eV, the variational 
cluster calculations (see Sec.~\ref{Sec:VCA} below) indicate the parent 
compound to be a Mott insulator. Therefore, to keep the system metallic 
in agreement with experiments, we were forced to choose a lower value of $U=2$~eV.

\section{Random Phase Approximation}\label{Sec:RPA}
For the case when interaction is not too strong, 
we can treat the Hubbard terms and Hund's term as perturbation, and do an RPA calculation 
for both the spin-spin correlation function and orbital nematic density-density correlation function:
\begin{align}
&\quad \chi_{spin} ( \bm{k}, i\omega_n ) \nonumber \\
& \equiv 2\int_0^\beta d\tau \sum_{\alpha_1 \alpha_2} 
\sum_{\bm{r}} e^{i \omega_n \tau -i \bm{k} \cdot \bm{r}} 
\langle T \hat{\bm{S}}^z_{\alpha_1} (\bm{r}, \tau) \hat{\bm{S}}^z_{\alpha_2} (\bm{0}, 0) \rangle
\end{align}
\begin{align}
&\quad \chi_{nematic} ( \bm{k}, i\omega_n ) \nonumber \\
& \equiv \frac{1}{2} \int_0^\beta d\tau \sum_{\bm{r}} e^{i \omega_n \tau -i \bm{k} \cdot \bm{r}} 
\langle T (\hat{n}_{xz} (\bm{r}, \tau)-\hat{n}_{yz} (\bm{r}, \tau))\, \cdot  \nonumber \\
&\quad \cdot (\hat{n}_{xz} (\bm{0}, 0)-\hat{n}_{yz} (\bm{0}, 0)) \rangle
\end{align}
Where $\alpha_1,\alpha_2$ are orbital indices. Summation over spin indices has been made implicit by writing 
$\hat{n}_\alpha = \hat{n}_{\alpha \uparrow}+\hat{n}_{\alpha \downarrow}$.

The bare spin and orbital nematic susceptibilities are of the form:
\begin{align}\label{Eq:chi_spin0}
&\quad \chi_{spin}^{(0)} (\bm{k}, i\omega_n ) \nonumber \\
&= \frac{-1}{2 \beta \mathcal{N}} \sum_{\omega_m , \bm{q}} 
\Tr [\sigma^z_{s_1 s_2} \tilde{G}_0^{\alpha_1 \alpha_2} (\bm{k}+\bm{q},i\omega_n +i \omega_m) 
    \sigma^z_{s_1 s_2} \tilde{G}_0^{\alpha_2 \alpha_1} (\bm{q},i\omega_m)] \nonumber \\
&=-\left(D_{0}^{xx}(\bm{k}, i\omega_n )+D_{0}^{yy}(\bm{k}, i\omega_n )
+2D_{0}^{xy}(\bm{k}, i\omega_n )\right)
\end{align}
\begin{align}\label{Eq:chi_nematic0}
&\quad \chi_{nematic}^{(0)}(\bm{k}, i\omega_n ) \nonumber \\
&= \frac{-1}{2 \beta \mathcal{N}} \sum_{\omega_m , \bm{q}} 
\Tr [\tau^z_{\alpha_1 \alpha_2} \tilde{G}_0^{\alpha_1 \alpha_4} 
    (\bm{k}+\bm{q},i\omega_n + i\omega_m) 
    \tau^z_{\alpha_3 \alpha_4} \tilde{G}_0^{\alpha_3 \alpha_2} 
    (\bm{q},i\omega_m)] \nonumber \\
&=-\left(D_{0}^{xx}(\bm{k}, i\omega_n )+D_{0}^{yy}(\bm{k}, 
i\omega_n )-2D_{0}^{xy}(\bm{k}, i\omega_n ) \right)
\end{align}
Where trace is over both spin and orbital indices. $\tilde{G}_0^{\alpha \beta}$ is the non-interacting Green's function. 
$D_{0}^{\alpha\beta}(\bm{k},i\omega_n)$ is the $2\times2$ matrix 
standing for the bare bubble, and $\alpha,\beta=x,y$ stand for $xz,yz$ orbitals:
\begin{minipage}{8cm}
\centering
%\flushleft
%\begin{figure}[H]
\includegraphics[scale=0.45]{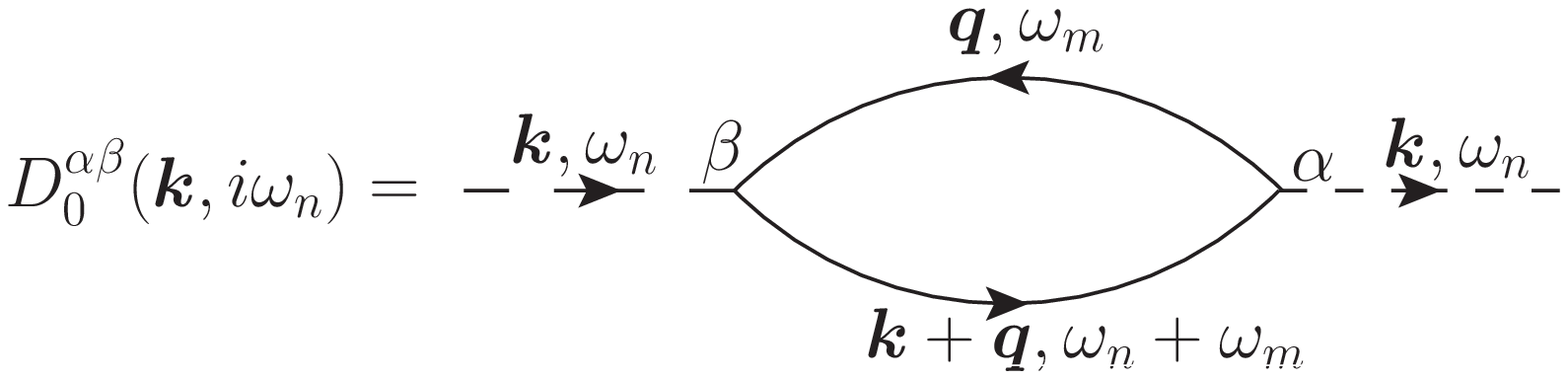} 
%\end{figure}
\begin{align}\label{Eq:D0}
&= \frac{1}{\mathcal{N}} \sum_{\bm{q}} \sum_{\nu_1 \nu_2} 
\langle \alpha | \nu_1,\bm{k}+\bm{q} \rangle 
\langle \nu_1,\bm{k}+\bm{q}|\beta \rangle \langle \beta | 
\nu_2, \bm{q} \rangle \langle \nu_2, \bm{q} | \alpha 
\rangle \nonumber \\
&\quad \cdot \frac{n_F (E_{\bm{q},\nu_2}) - n_F (E_{\bm{k}+\bm{q},\nu_1})}{i \omega_n 
+ E_{\bm{q},\nu_2}- E_{\bm{k}+\bm{q},\nu_1}}
\end{align}
\end{minipage}
\vspace{3mm}

%&&D_0^{\alpha \beta}(\bm{k}, i\omega_n ) \nonumber \\*
\noindent
where $\nu_1, \nu_2$ are band indices, and $E_{\nu,\bm{q}}$ is the dispersion 
of the two non-interacting bands.\cite{Raghu08}
\begin{equation}
E_\pm (\bm{k}) = \epsilon_+ (\bm{k}) \pm \sqrt{\epsilon^2_- (\bm{k})+ \epsilon_{xy}^2 (\bm{k})}
\end{equation}

The matrix form of the bare spin and orbital nematic susceptibilities are
as follows: 
\begin{equation}
\big( \chi_{spin}^{(0)} \big) = 
\left( 
\begin{array}{cc}
\chi_{xx}^{(0)} & \chi_{xy}^{(0)} \\
\chi_{yx}^{(0)} & \chi_{yy}^{(0)}
\end{array}
\right)_{spin}=-
\left( 
\begin{array}{cc}
D_0^{xx} & D_0^{xy} \\
D_0^{yx} & D_0^{yy}
\end{array}
\right)
\end{equation}
\begin{equation}
\big( \chi_{nematic}^{(0)} \big) = 
\left( 
\begin{array}{cc}
\chi_{xx}^{(0)} & \chi_{xy}^{(0)} \\
\chi_{yx}^{(0)} & \chi_{yy}^{(0)}
\end{array}
\right)_{nematic}=-
\left( 
\begin{array}{cc}
D_0^{xx} & -D_0^{xy} \\
-D_0^{yx} & D_0^{yy}
\end{array}
\right)
\end{equation}

\begin{figure}[!htbp]
\includegraphics[scale=0.4]{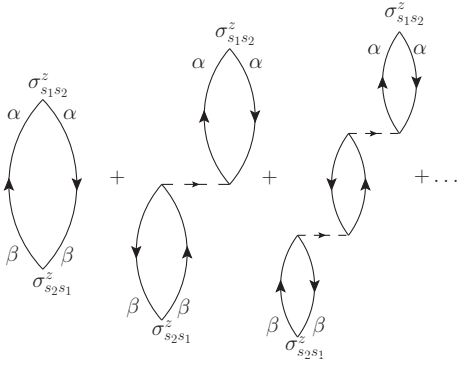}
\caption{\label{fig:RPA-spin}RPA diagrams for spin susceptibility.}
\end{figure}

\begin{figure}[!htbp]
\includegraphics[scale=0.4]{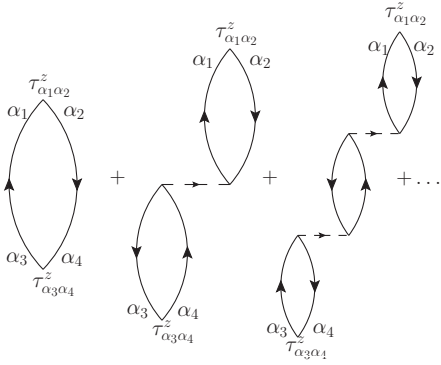}
\caption{\label{fig:RPA-nematic}RPA diagrams for orbital nematic susceptibility.}
\end{figure}

When taking interactions into consideration, we sum over the RPA series 
of diagrams for the two susceptibilities, 
see Figs.~\ref{fig:RPA-spin} and ~\ref{fig:RPA-nematic}.

Then RPA-renormalized spin and orbital susceptibilities take the form:
\begin{equation}
\big( \chi_{spin} \big) = \big( \chi_{spin}^{(0)} \big) \left( \mathbb{1} -
\left(
\begin{array}{cc}
U & J \\
J & U
\end{array}
\right)
\big( \chi_{spin}^{(0)} \big) \right)^{-1}
\end{equation}
\begin{equation}
\big( \chi_{nematic} \big) = \big( \chi_{nematic}^{(0)} \big) \left( \mathbb{1} +
\left(
\begin{array}{cc}
U & -2U^\prime +J \\
-2U^\prime +J & U
\end{array}
\right)
\big( \chi_{nematic}^{(0)} \big) \right)^{-1}
\end{equation}


When we sum over all components of the matrices, we get the total 
RPA-renormalized susceptibilities in the scalar form:
%\begin{align}
\begin{eqnarray} 
 & & \chi_{spin} (\bm{k}, i\omega_n ) \nonumber \\
 & =&  \chi_{xx}(\bm{k}, i\omega_n ) + \chi_{xy}(\bm{k}, i\omega_n ) +\chi_{yx}(\bm{k}, i\omega_n ) +\chi_{yy}(\bm{k}, i\omega_n )  \nonumber \\*
 & =&  \Big[\chi_{spin}^{(0)} (\bm{k}, i\omega_n )-2(U-J)\det D_{0}(\bm{k}, i\omega_n )\Big]/\Big\{1-U \chi_{spin}^{(0)} (\bm{k}, i\omega_n ) \nonumber \\
 & \quad & -2(U-J)D_{0}^{xy}(\bm{k}, i\omega_n ) +  (U^{2}-J^{2})\, \det D_{0}(\bm{k}, i\omega_n )\Big\}
%\end{align}
\end{eqnarray}
\begin{align}
 & \quad \chi_{nematic}(\bm{k}, i\omega_n )\nonumber \\
 & =  \Big[\chi_{nematic}^{(0)}(\bm{k}, i\omega_n )+2(U+2U^{\prime}-J)\det D_{0}(\bm{k}, i\omega_n )\Big]/\Big\{1+\nonumber \\*
 & \quad U\chi_{nematic}^{(0)}(\bm{k}, i\omega_n )+2(U-2U^{\prime}+J)D_{0}^{xy}(\bm{k}, i\omega_n ) +\nonumber \\*
 & \quad [U^{2}-(2U^{\prime}-J)^{2}]\, \det D_{0}(\bm{k}, i\omega_n )\Big\}
\end{align}
Where $\chi_{spin}^{(0)}$ and $\chi_{nematic}^{(0)}$ are defined in Eq.~\ref{Eq:chi_spin0} and Eq.~\ref{Eq:chi_nematic0}, 
and $\det D_{0}=D_{0}^{xx} D_{0}^{yy}-D_{0}^{xy} D_{0}^{yx}$.

From the expression of $D_0^{\alpha \beta}$ (see Eq.~\ref{Eq:D0}), 
Re$D_0^{\alpha \beta}<0$ in general. We also notice that at 
$\bm{q}=0$, $\det D_{0}(\bm{0}, \omega )>0$, thus the ferro orbital 
nematic instability comes from the $-(2 U^\prime-J)^2 \, \det D_0$ term, 
ie. the orbital nematic susceptibility at $\bm{q}=0$ only diverges when 
we have large enough inter-orbital Hubbard repulsion, and Hund's coupling 
cannot be too large.

\begin{figure}[!tbp]
\includegraphics[width=0.55\textwidth]{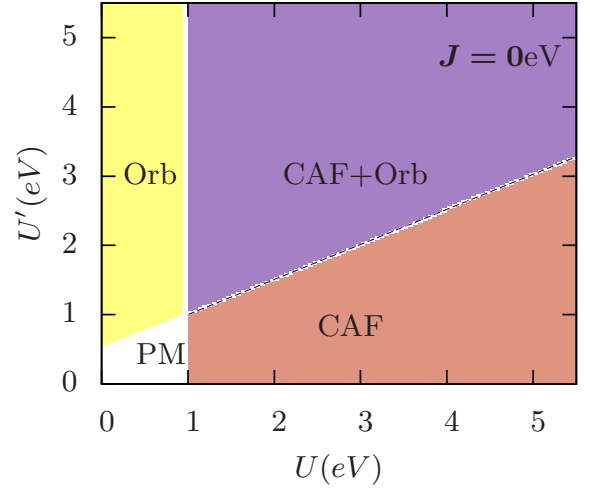}
\caption{\label{fig:RPA-phd3}(Color online) RPA phase diagram when fixing $J=0$ and 
$\mu=0.3625 \text{eV}$. 
Phases PM, CAF and Orb respectively denote the isotropic paramagnetic phase, 
the columnar antiferromagnetic phase, and the orbital nematic phase.}
\end{figure}
\begin{figure}[!tbp]
\includegraphics[width=0.55\textwidth]{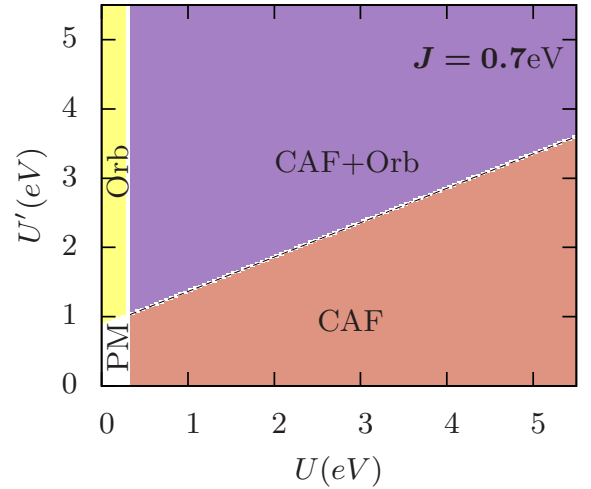}
\caption{\label{fig:RPA-phd4}(Color online) RPA phase diagram when fixing $J=0.7eV$ 
and $\mu=0.3625 \text{eV}$. 
Phases PM, CAF and Orb using same abbreviation as in Fig.~\ref{fig:RPA-phd3}}
\end{figure}
At small interaction strength, the spin and orbital susceptibilities are finite. 
Upon increasing interaction, spin susceptibility at $\mathbf{Q}=(\pi,0)$ or $(0,\pi)$ and 
orbital nematic susceptibility at $\mathbf{q}=(0,0)$ start 
to diverge in a certain parameter region, implying the tendency towards
columnar antiferromagnetic order and ferro orbital 
nematic order, respectively. The phase boundaries are thus given when the RPA 
renormalized susceptibilities diverge. 

\begin{figure}[!tbp]
\includegraphics[scale=1]{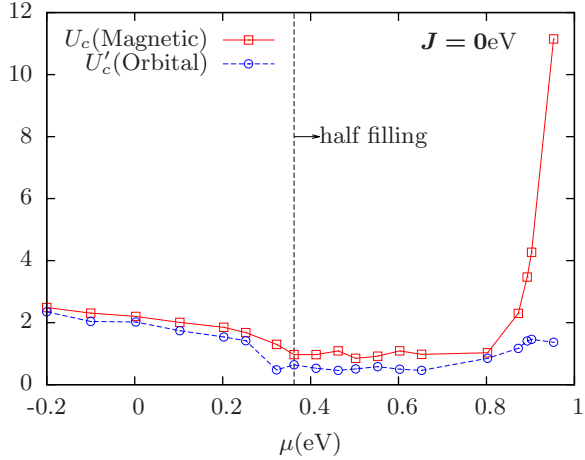}
\caption{\label{fig:RPA-critical-J0d0}(Color online) The dependence of critical values 
of interactions on $\mu$ in the absence of Hund's coupling, $J=0$.}
\end{figure}
\begin{figure}[!tbp]
\includegraphics[scale=1]{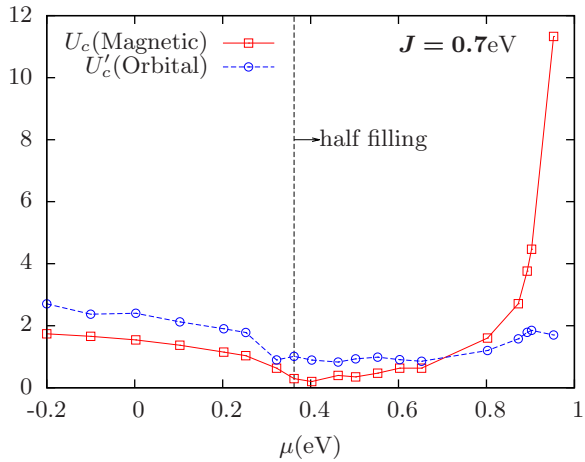}
\caption{\label{fig:RPA-critical-J0d7}(Color online) The dependence of critical values 
of interactions on $\mu$ for finite Hund's coupling $J=0.7$~eV.}
\end{figure}

In Fig.~\ref{fig:RPA-phd3} and Fig.~\ref{fig:RPA-phd4}, we 
set Hund's coupling $J=0$ and $J=0.7eV$, respectively. In both phase diagrams, 
we observe four separate regions: the isotropic paramagnetic phase (PM), 
the columnar antiferromagnetic phase (CAF), 
the orbital nematic phase (Orb), and a region where both susceptibilities 
have diverged (CAF+Orb). In the CAF region, since magnetic order has already 
developed ($\langle M_1\rangle \neq \langle M_2\rangle $), the orbital nematic order should also be nonzero since there is 
a linear coupling between orbital order and magnetism, see Eq.~(\ref{Eq.free}).
%of the form $\eta \langle p \rangle \cdot \left\langle M_1^2 - M_2^2 \right\rangle$, as mentioned earlier in Sec.~\ref{Sec:Intro}.  
We therefore used dashed line in the phase diagrams to indicate that the CAF and CAF+Orb phases are 
actually not distinguishable. This is not true for the boundary between 
Orb and CAF+Orb phases however, since finite orbital order ($\langle p\rangle\neq 0$) does not necessarily lead to long-range magnetic order: magnetic fluctuations can break the $C_4$ symmetry, $\langle\psi\rangle\neq 0$ in Eq.~(\ref{Eq.psi}), without a true magnetic order\cite{CCL}.
%through the $\eta \langle p \rangle \cdot  \left(\langle M_1^2 \rangle - \langle M_2^2 \rangle \right)$ coupling, even when there is finite orbital order, there can be magnetic fluctuations without true magnetic order.

%In Fig., we set Hund's coupling $J=0.7eV$. \blue{The line of requirement 
%$U^\prime=U-2J$ passes through phase II and II,III only.}

%The phase diagram could be helped understood in the atomic limit of two rigid bands. 
%Assuming there is a small splitting $\Delta$ of the two bands with band width $W$, 
%the energy gain of the splitting will be approximately:
%\begin{displaymath}
%E=U\frac{| \Delta |}{W}+J(\frac{| \Delta |}{W}-1)-U^\prime(\frac{\Delta^2}{W^2}-1)
%\end{displaymath}
%Thus inter-orbital $U^\prime$ term will favor the splitting of the two bands while intra-orbital $U$ and Hund's coupling $J$ not. 
%More detailed analysis of the splitting energy will be discussed in Sec.~\ref{Sec:mf}. In Eq.~\ref{Eq:freeEnergy-p}, if we expand 
%the $\log$ assuming small orbital nematic order, we will naturally get higer order terms including $\Delta^4$.
\begin{figure}[!tbp]
\includegraphics[scale=1]{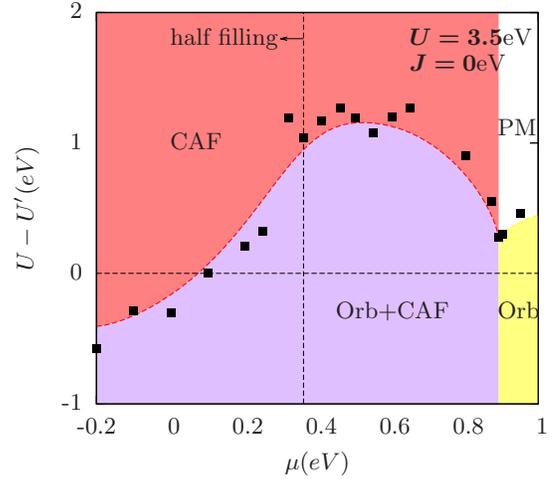}
\caption{\label{fig:RPA-phd-doping-J0d0}(Color online) RPA phase diagram 
when fixing $U=3.5$eV, $J=0$. Notice that for sufficiently high electron doping, the orbital nematic phase without magnetic instability (yellow region) appears even when intra-orbital interaction dominates $U>U^\prime$. Dashed vertical line marks the half-filled case (parent compound).}
\end{figure}
\begin{figure}[!tbp]
\includegraphics[scale=1]{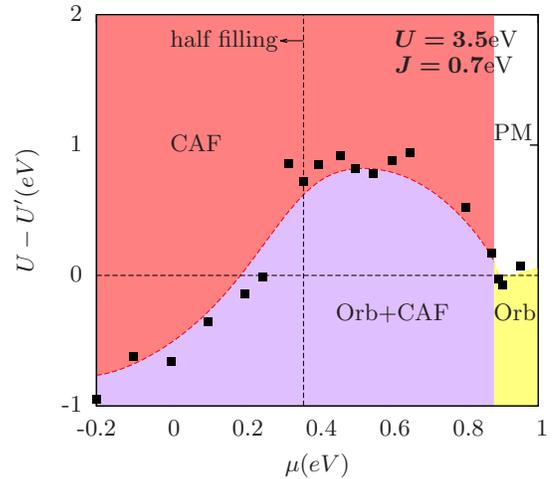}
\caption{\label{fig:RPA-phd-doping-J0d7}(Color online) RPA phase diagram 
when fixing $U=3.5$eV, $J=0.7$eV. As in Fig.~\ref{fig:RPA-phd-doping-J0d0}, a non-magnetic orbital nematic phase (yellow region) is stabilized by electron doping. Dashed vertical line marks the half-filled case (parent compound).}
\end{figure}

We have also studied the effect of doping on the phase diagram. Let's denote 
the critical value of $U$ when the system enters the magnetic CAF phase as $U_c$ (while 
keeping $U^\prime=0$), and denote the critical value of $U^\prime$ when the 
system becomes orbitally ordered as $U^\prime_c$ (while keeping $U=0$). The dependence 
 of $U_c$ and $U_c^\prime$ on doping is plotted in Fig.~\ref{fig:RPA-critical-J0d0} for zero Hund's coupling and in Fig.~\ref{fig:RPA-critical-J0d7} for a realistic value of $J=0.7$~eV.

From Figs.~\ref{fig:RPA-critical-J0d0} and \ref{fig:RPA-critical-J0d7} 
we see that both electron and hole doping enhance the critical value of 
interactions necessary to stabilize antiferromagnetic or orbital order, making it more difficult to enter the ordered phases. However, 
the doping effect is not particle-hole symmetric: electron doping 
greatly enlarges $U_c$, which means that magnetic order is suppressed 
much faster than orbital nematic order upon electron doping.
To make this more transparent, we plotted the phase diagrams describing both the effects of interactions and doping 
in Fig.~\ref{fig:RPA-phd-doping-J0d0} and Fig.~\ref{fig:RPA-phd-doping-J0d7}. 
Besides the effect that larger $U^\prime$ makes the orbital nematic 
phase more stable, which is consistent with Fig.~\ref{fig:RPA-phd3} and 
Fig.~\ref{fig:RPA-phd4}, 
we also notice that the phases are very sensitive to doping. In particular, 
when the system is sufficiently doped with electrons, we found that the magnetic susceptibility 
is always finite, so that as a function of increasing $U'-U$, the only phase transition is from paramagnetic phase 
to the orbital nematic phase, without any magnetism. Interestingly, the orbital ordered phase can be stabilized even when the intra-orbital repulsion $U$ dominates ($U>U^\prime$), see Fig.~\ref{fig:RPA-phd-doping-J0d0}. Normally, the regime $U>U^\prime$ is expected to be dominated by antiferromagnetism (c.f. the half-filled case, marked by a dashed vertical line), however in the case of large electron doping the propensity to magnetic ordering is strongly suppressed, resulting in a non-magnetic orbital nematic phase (yellow region in Figs.~\ref{fig:RPA-phd-doping-J0d0} and Fig.~\ref{fig:RPA-phd-doping-J0d7}).

\section{Variational Cluster Approximation}\label{Sec:VCA}
The RPA is a weak-coupling approach that only detects the tendency
to certain orderings based on the divergence of the respective susceptibilities. To study the ordered phases themselves, we use the 
variational cluster approximation 
(VCA)\cite{Potthoff03}, which is a non-perturbative quantum cluster 
method similar in spirit to the cluster dynamical mean-field theory (CDMFT)\cite{Kotliar01}. 
Unlike in the CDMFT however, the bath degrees of freedom are not included 
explicitly in the calculation. Rather, the effect of the bath is captured 
indirectly by varying the inter-cluster one-body parameters $\{h\}$ in such 
a way as to minimize the free energy (Potthoff functional) $\Omega[\Sigma_{ij}(\omega)]$ 
calculated in the conserving approximation~\cite{Potthoff03}. The Potthoff 
functional depends on the cluster self-energy $\Sigma_{ij}(\omega,\{h\})$, 
which in turn depends on the variational parameters of the cluster $\{h\}$. 
These parameters are fixed from the variational principle on the Potthoff 
functional: $\delta \Omega[\Sigma]/\delta \Sigma=0$ at the solution. 
In practical calculations, the variational principle is enforced by 
requiring that $\partial \Omega[\Sigma(\{h\})]/\partial\{h\}=0$.

Formulated in this fashion, the VCA is a variational extension of the 
cluster perturbation theory\cite{Gros93,Senechal2000} and provides a powerful way of 
treating the strongly correlated lattice models with local (on-site) 
interactions. The VCA method has been shown to capture both the weak- 
and strong-coupling limit of the (one-band) Hubbard model and compares 
very favourably to the quantum Monter Carlo simulations.~\cite{}
It has been used successfully to study the metal-insulator transition~\cite{Balzer2008, Balzer2009}, 
frustrated magnetism~\cite{Nevidomskyy2008, Sahebsara2008} and d-wave 
superconductivity in quasi-2D organic superconductors \cite{Sahebsara2006} 
and in the high-Tc cuprates \cite{Senechal2004, Senechal2005, Nevidomskyy2008}. It was 
shown in particular to capture the d-wave superconductivity of purely 
electronic origin and to yield the correct doping dependence\cite{Senechal2005}, 
as well as the pseudogap feature in the quasiparticle spectral weight\cite{Senechal2004, Tremblay2006}.
 We use the exact diagonalization (ED) method based on Lanczos method 
 to solve the quantum cluster impurity model, which offers two principal 
 advantages over the Monte-Carlo based solvers routinely used in CDMFT: 
 (i) there is no need for analytical continuation as the self-energy is 
 expressed in real, not imaginary, frequency and (ii) the zero-temperature
properties can be readily accessed, avoiding the infamous
fermionic sign problem inherent to the quantum Monte Carlo impurity solvers.

The VCA method is particularly well suited to our task because it allows 
explicit treatment of a spontaneous symmetry-breaking long-range order, 
and has been successfully used to study magnetism~\cite{Dahnken2004, Nevidomskyy2008, Sahebsara2008} 
and unconventional superconductivity~\cite{Senechal2004,Senechal2005,Nevidomskyy2008} in the Hubbard model. 
To study the orbital nematic order, we 
allow the Hamiltonian on a cluster to have
a variational degree of freedom associated with a cluster Weiss field,  
$\Delta \hat{H}_{cl} = p_\text{cl} \cdot (\hat{n}_{xz}-\hat{n}_{yz})$. Note 
that the actual lattice Hamiltonian is unaltered, so that the $C_4$ symmetry 
is not explicitly broken by construction. Rather, the spontaneously broken symmetry inside the orbital nematic phase is signified by a non-zero value of the cluster Weiss field 
$p_\text{cl}$ at the variational solution of the Potthoff functional~\cite{Potthoff03}:
\begin{equation}
\left.\frac{\delta \Omega[\Sigma(p_\text{cl})]}{\delta p_\text{cl}}\right|_\text{sol} =0.
\label{eq:saddle}
\end{equation}
This is in difference to earlier VCA calculations of nematicity by Daghofer 
and collaborators~\cite{Daghofer12a, Daghofer12b}, in which the $C_4$ 
tetragonal symmetry was  broken by construction  at the level of the 
original lattice Hamiltonian, by introducing anisotropy either in the 
onsite energy of the $xz/yz$ orbitals, in the hopping amplitudes between 
the $x$ and $y$ directions, or in the Heisenberg exchange terms in 
the $x$ and $y$ directions. 
%We find such treatment unsatisfactory because the nematic order had 
%been imposed ``by hand,'' rather than emerging as a spontaneously broken symmetry.
These studies do not  probe the spontaneous symmetry breaking but rather investigate the response of 
the system to the $C_2$ distortion, similar in spirit to the experimental 
studies under uniaxial stress~\cite{Chu10} or strain~\cite{Chu12}. 
The present approach, on the other hand, is faithful to the original 
variational idea by Potthoff~\cite{Potthoff03}, allowing the  $C_4$ 
symmetry to be broken spontaneously, by introducing the in-cluster 
nematic Weiss field $p_\text{cl}$.
\begin{figure}[!tbp]
\includegraphics[scale=1]{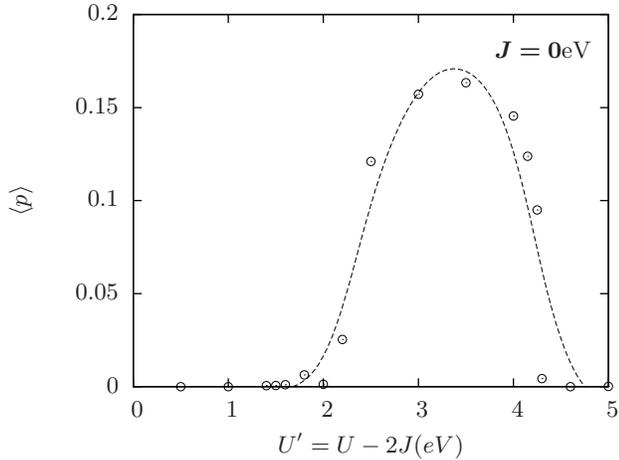}
\caption{\label{fig:vca-p-up-J0}Orbital nematic order parameter versus 
interaction strength from VCA calculation at half-filling. The inter-orbital Hubbard repulsion $U^\prime$ is varied at the same time as the intra-orbital $U$: $U^\prime = U-2J$, with the 
Hund's coupling $J$ fixed at zero. The dotted line is a guide to the eye.}
\end{figure}
\begin{figure}[!tbp]
\includegraphics[scale=1]{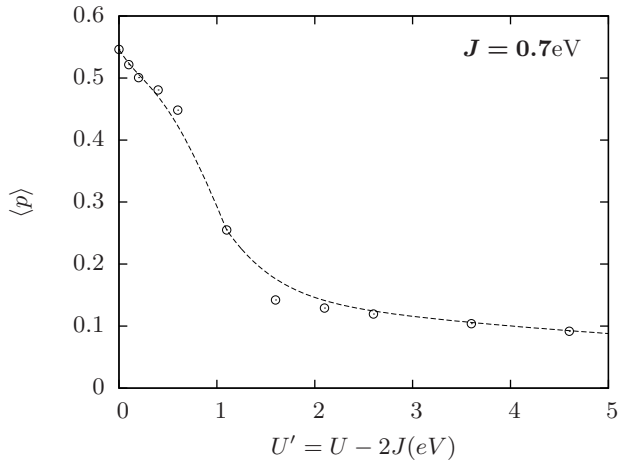}
\caption{\label{fig:vca-p-up-J0d28}Orbital nematic order parameter versus the inter-orbital Hubbard coupling $U^\prime = U-2J$ from VCA calculation at half-filling. 
Hund's coupling is fixed at $J=0.7eV$. The dotted line is a guide to the eye.}
\end{figure}

\begin{figure}[!tbp]
\includegraphics[scale=1]{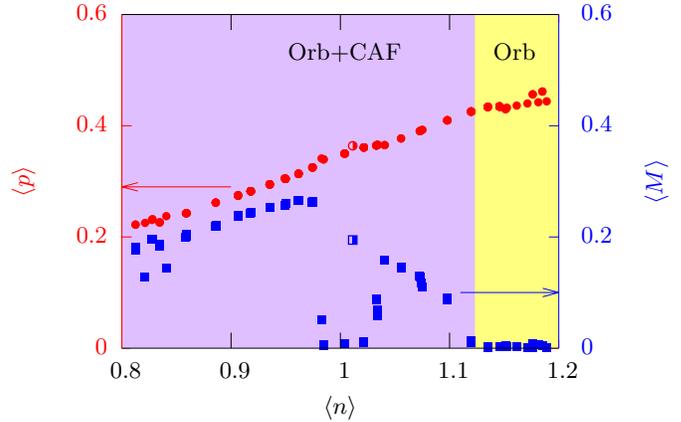}
\caption{\label{fig:vca-pM-doping}(Color online) Change of orbital nematic order (circle) and 
magnetic order(square) with respect to doping from VCA calculation, when $U=2\text{ eV}$, 
$U^{\prime}=0.6\text{ eV}$, $J=0.7\text{ eV}$. 
The half-full symbol corresponds to not a smooth minima but a cusp in the Potthoff functional.  Note that near half filling, solutions with zero staggered magnetization appear -- these are unphysical and should be disregarded due to the VCA minimization algorithm becoming unstable when the chemical potential falls onto sharp maxima/minima in the density 
of states.% in the simplified two-orbital model.
This does not however affect any of our conclusions.
}
\end{figure}

From the solution of Eq.~(\ref{eq:saddle}), we then obtain the cluster 
propagator and self-energy using the ED solver. The full lattice propagator 
$G(\mathbf{K},\omega)$ is then calculated from the cluster solution by 
treating the one-body hopping terms between neighboring clusters as a perturbation.
When the Potthoff functional is minimized at a nonzero value of $p_\text{cl}$, the 
full lattice Hamiltonian will develop a long range order of the orbital nematicity, 
ie. a nonzero value of $\langle p \rangle \equiv \langle \hat{n}_{xy}-\hat{n}_{yz} \rangle$.

In Fig.~\ref{fig:vca-p-up-J0}, the expectation value of the orbital nematic order 
parameter is calculated on the lattice, as a function of $U^{\prime}=U$. 
After developing a non-zero value of $\langle p \rangle$, 
upon further increasing $U^\prime = U$, the orbital nematic order parameter 
attains a peak and then becomes suppressed at higher values of the interaction 
strength.
In the calculation for a realistic value of Hund's coupling $J=0.7$~eV (Fig.~\ref{fig:vca-p-up-J0d28}), 
the magnitude of order parameter $p$ starts from a non-zero value when $U^{\prime}$ 
is small. This is somewhat unexpected from our RPA results, 
where Hund's coupling $J$ slightly suppresses orbital nematic order (see Fig.~\ref{fig:RPA-phd3}, 
\ref{fig:RPA-phd4}, \ref{fig:RPA-critical-J0d0} and \ref{fig:RPA-critical-J0d7}). 
Upon further increasing $U^\prime$, the same suppression of $p$ is observed  as in the case of $J=0$ above (see 
 Fig.~\ref{fig:vca-p-up-J0}).

%\begin{figure}[!htbp]
%\includegraphics[scale=1]{energy_den_vca.eps}
%\caption{\label{fig:energy-density}Ground state energies of two solutions. 
%The higher curve corresponds to hollow data points in Fig.~\ref{fig:vca-pM-doping}
%}
%\end{figure}

The above VCA calculations have probed the orbital nematic order in the absence of antiferromagnetism. Of course we know from our RPA calculations that magnetism also arises in the phase diagram of the two-orbital model.
The effects of interaction on magnetic order alone (without orbital order) have been studied with VCA in Ref.~\onlinecite{Daghofer08}. The authors found that the magnetic order starts 
developing at intermediate interactions $U$ in the parent compounds. (Although it should be noted that in their study, the Hund's coupling $J$ was assumed to scale with $U$ as $J=U/4$, which becomes unphysical for too small or too large values of $U$). 
  
We will now investigate the phase diagram of the model as a function of electron doping when both the orbital order and columnar antiferromagnetism are present. We set the interaction strengths close to realistic values determined from the \emph{ab initio} constrained-RPA calculations\cite{Aichhorn09}:
 $U=2$~eV, $J=0.7$~eV, $U^\prime=U-2J=0.6$~eV. 
Besides the orbital nematic Weiss field $p_\text{cl}$ coupled to $n_{xy}-n_{yz}$ 
on the cluster, we have also introduced the columnar 
antiferromagnetic Weiss field $M_\text{cl}$ coupled to $S^z_{\bm{r}} \cos 
(\bm{Q} \cdot \bm{r})$ on the cluster, with wave vector 
$\bm{Q}=(\pi,0)$ or $(0,\pi)$. A VCA variational 
search was then performed for both $p_\text{cl}$ and $M_\text{cl}$, to study the interplay of the 
two orders. 
The resulting doping dependence is plotted  in Fig.~\ref{fig:vca-pM-doping},
where the lattice chemical potential was varied to control the electron occupancy.
%we fixed the orbital occupation  on the cluster to be one per site 
%and varied the lattice chemical potential
%in order to study the doping dependence of the magnetic and orbital nematic order parameters. 
We found coexisting orbital and antiferromagnetic orders 
in the parent compound at half-filling. While both types of ordering are suppressed by hole doping, the effect of electron doping is to slightly enhance the orbital nematic order while suppressing antiferromagnetism. This  effect likely stems from the competition between the two orders, although the enhancement of orbital order with electron doping could be an artifact of the simplified two-orbital model studied here.

Crucially, we find that upon electron doping, antiferromagnetism
is suppressed much faster than orbital nematic phase, resulting in
a region at sufficiently high electron doping ($x\gtrsim13\%$) where orbital nematic order exists without any magnetism. This is consistent with our previous RPA finding (yellow region in 
Figs.~\ref{fig:RPA-phd-doping-J0d0} and \ref{fig:RPA-phd-doping-J0d7}).

%The full(and half-full) symbols represent 
%one set of solution with relatively lower found state energies, while hollow 
%symbols represent another set of solution with higher energies(Fig.~\ref{fig:energy-density}).

%As a comparison,  we have also plotted a higher-energy solution (hollow symbols), 
%with zero orbital nematic order but large magnetic order (so large that it opened 
%up a gap in the electron bands and made the parent compound an insulator). 
%The  hollow points also represent local minimas as a solution of the problem, but with relatively higher energies (see Fig.~\ref{fig:energy-density}).

\section{Discussion}\label{Sec:Con}
We have studied the emergence of orbital nematic order in the iron pnictide 
superconductors, within the framework of the two-orbital Hubbard model that 
captures the physics of Fe $d_{xz}$ and $d_{yz}$ orbitals. In particular, 
we have analyzed the dependence of the nematic order on doping and interaction 
strength, as well as its interplay with magnetism.  

First,  we studied the instabilities towards the orbital 
nematic order and magnetic order in the weak-coupling approach by calculating 
the corresponding susceptibilities using the RPA method. 
We recovered the results of previous studies that the ordering wave-vector is at 
$\mathbf{Q}=(\pi,0)$ or $(0,\pi)$ for columnar antiferromagnetism\cite{Raghu08} 
and $(0,0)$ for ferro-orbital nematic order\cite{Lee09-LDA+U}, respectively. 
We found that orbital nematic order strongly depends on inter-orbital 
Hubbard repulsion $U^{\prime}$, while magnetic order depends on the 
intra-orbital Hubbard repulsion $U$. Both the magnetic and orbital 
nematic order parameters are affected by Hund's coupling $J$: larger 
$J$ values tend to suppress the  propensity to orbital nematic ordering and, on the other hand, enhance 
the magnetic susceptibility.

It has been long believed that the nematicity and antiferromagnetism coexist at low temperatures, with a wealth of experimental data supporting this in the 1111 and 122 families of iron pnictides. Within the RPA, we indeed find regions in the phase space where both the magnetic 
and nematic susceptibilities diverge, implying coexistence of the two order 
parameters.
It is true that the long-range CAF phase with $M_1\neq M_2$ (the notation introduced in Sec.~\ref{Sec:Intro}) necessarily breaks the $C_4$ symmetry and will generically induce a non-zero value of orbital polarization $p$ because of the linear coupling in the Landau free energy. The converse is however not true: the ferro-orbital phase with non-vanishing $p$, while nematic in nature, need not have long-range magnetic order. Indeed, our RPA calculations show that upon electron 
doping, magnetic susceptibility is suppressed much faster than orbital ordering, until for sufficiently large electron doping, only orbital order survives
(See Fig.~\ref{fig:RPA-critical-J0d0} and Fig.~\ref{fig:RPA-critical-J0d7}). Hole doping, on the other hand, does not reveal this tendency.

We note that our results do not eliminate the possibility of spin fluctuations taking part in the nematicity even if the static magnetic order is absent. Indeed, as remarked in the Introduction, ferro-orbital nematic order parameter will couple linearly to the spin-fluctuation Ising parameter, see Eq.~(\ref{Eq.free}). However, when the system is sufficiently far from the antiferromagnetic phase, as is arguably the case in FeSe\cite{McQueen09, Bendele10}, spin fluctuations are expected to be weak, whereas a non-zero static ferro-orbital order will be the main driver of the nematicity.  

Since RPA is a weak coupling approach which works only when the long-range order is approached from the disordered phase,  we have used the non-perturbative variational cluster approximation (VCA) which allowed us to study 
the ordered phases in the variational approach.
We studied the orbital nematic and magnetic order parameters, as well as their interplay as a function of
electron density (doping) and interaction strength. 
Our VCA calculation showed that orbital nematic order strongly 
depends on inter-orbital Hubbard interaction $U^{\prime}$ and Hund's 
coupling $J$, similar to the RPA results. However, while Hund's 
coupling suppresses orbital order within the RPA approach, 
this is not the case in the non-perturbative VCA calculation. In Fig.~\ref{fig:vca-p-up-J0d28}, 
we find a non-vanishing orbital nematic order even in the absence of inter-orbital interaction $U^{\prime}=0$ and finite Hund's coupling $J=0.7$~eV, implying that the effect of interactions and Hund's term is not always captured properly by the RPA method. 

The doping dependence study from VCA shows that electron and 
hole doping are not symmetric. Crucially, we find that upon moderate electron doping ($\gtrsim13$\%, see Fig.~\ref{fig:vca-pM-doping}), long-range magnetic order is completely suppressed, while the orbital nematic order persists, similar to our RPA results.

It is instructive to compare our VCA results with previous attempts to address 
nematicity with the cluster dynamical mean-field theory  (CDMFT)\cite{Kotliar01} and similar 
cluster approaches. One possible way to detect tendency to nematicity is to 
explicitly introduce orthorhombic distortion into the hopping terms in 
Eqs.~(\ref{eq:H0}, \ref{eq:eps_k}), and then study the electronic response. 
For the one-band Hubbard model, this has been done using CDMFT\cite{Okamoto10} and dynamical cluster 
approximation\cite{Su11}. Both groups 
found that for a sufficiently large interaction $U$, a small orthorhombic 
distortion can lead to a large nematic response in the low-energy electron 
scattering rate. 
Another way to study the spontaneous development of nematicity in the one-band 
Hubbard model is by introducing an anisotropic hopping inside the cluster alone $\Delta \hat{H}_{cl}=
\delta t \sum_{\bm{r}}  (\hat{c}^\dagger_{\bm{r}} \hat{c}_{\bm{r}+x}-
\hat{c}^\dagger_{\bm{r}} \hat{c}_{\bm{r}+y})+h.c.$, and then optimize the strength of $\delta t$ variationally in VCA\cite{Fang13}. 
Using this approach, the authors found that anisotropy can develop in the overdoped region.\cite{Fang13}
However, the multi-orbital nature of the iron pnictides 
was not taken into account in these studies. 

The nematicity in multi-band 
Hubbard model has been studied with VCA in Refs.~\onlinecite{Daghofer12a, 
Daghofer12b}, however these authors have also explicitly broken the $C_4$ 
tetragonal symmetry at the level of the original lattice Hamiltonian, by 
introducing anisotropy either in the onsite energy of the $xz/yz$ orbitals, 
in the hopping amplitudes between the $x$ and $y$ directions, or 
in the Heisenberg exchange terms in the $x$ and $y$ directions.  Because 
the $C_4$ symmetry is broken by construction, these studies do not  probe 
the spontaneous symmetry breaking but rather investigate the response of 
the system to the $C_2$ distortion, similar in spirit to the experimental 
studies under uniaxial stress~\cite{Chu10} or strain~\cite{Chu12}. 
Not surprisingly, the magnitude of the induced orbital order is then proportional 
to the imposed strain and does not have an intrinsic value. 
In the present work, by contrast, the lattice expectation value of orbital 
order $p = \langle \hat{n}_{xz} -  \hat{n}_{yz}\rangle$ is finite even at 
zero induced strain and is consistent with the value found by ARPES in 
BaFe$_2$As$_2$ [\onlinecite{Yi11}].

We note that in a different context, nematicity in a two-orbital Hubbard model has also been studied in application to Sr${}_3$Ru${}_2$O${}_7$ in Refs.~\onlinecite{Tsuchiizu2013,Ohno2013}. In these works, 
the authors studied the nematic instability using RPA and renormalization group methods. It was found that Aslamazov--Larkin-type vertex corrections result in the 
strong coupling between spin and orbital fluctuations, leading the authors to conclude 
that the spin fluctuations lie at the origin of the nematic instability. 
In the present work, on the other hand, we find a regime of parameters where the ferro-orbital 
RPA susceptibility diverges even without the vertex corrections, while the spin susceptibility remains finite (see section~\ref{Sec:RPA}).
This implies that the spin fluctuations are not the primary origin of ferro-orbital nematicity in our model. Furthermore, from our VCA calculations 
we find that the orbital nematic order persists even when magnetic ordering is fully suppressed  in the electron doped region (see Fig.~\ref{fig:vca-pM-doping}), 
which again suggests that magnetic fluctuations are not the origin of orbital nematicity in 
our case. 
It should be noted that the Fermi surface topology and nesting properties in 
Sr${}_3$Ru${}_2$O${}_7$ are different from the iron pnictices, so the conclusions drawn in Refs.~\onlinecite{Tsuchiizu2013}, \onlinecite{Ohno2013} do not trivially generalize to our case.

 Finally, we note that other types of orbital ordering, involving $d_{xy}$ orbitals, 
 have been proposed in the literature,\cite{Ying11, Kontani11,Onari2012} however those are beyond the scope of the two-orbital model used in this study. While it would be desirable to extend this study to include all five iron orbitals, unfortunately the VCA calculations become computationally prohibitive because of the limitations of the exact diagonalization solver when dealing with multiple orbitals. Nevertheless, as remarked earlier in Sec.~\ref{Sec:Model}, we hope  that the present two-orbital model captures  the salient features of nematicity in the iron pnictides, based on the dominant contribution of $d_{xz}$ and $d_{yz}$ orbitals to the Fermi surfaces of these materials.\cite{Graser09}

\section{Conclusions}\label{Sec:Con2}
To summarize,  we have studied the doping dependence of both the 
orbital nematic and antiferromagnetic orders using the RPA and 
non-perturbative variational cluster approximation, and found a region at moderately large electron 
doping where the orbital nematic order survives without long-range magnetism. 
While these results are limited to the two-orbital model, which is not sufficient to describe the realistic band structure of the iron-based superconductors, 
our findings are suggestive of the connection to the experimental observations 
of an orbital nematic phase in FeSe\cite{Shimojima14-FeSe, Baek14, Boehmer14} without any 
sign of antiferromagnetism.\cite{McQueen09, Bendele10} 
This raises the question whether two-fold symmetric antiferromagnetic fluctuations are essential 
for stabilizing the nematic phase, as has been argued previously within 
the spin-nematic scenario.\cite{Dai09, Fernandes11, Fernandes12-sst, 
Fernandes12-prl, Fernandes14} It would appear from the present study that this may not always be the case, and while the importance of spin fluctuations is undeniable in the vicinity of the magnetic order in the 122 and 1111 families of iron pnictides, it is the orbital fluctuations that appear to be critical for the onset of nematicity in FeSe.
This situation may be more common than previously appreciated: recent ARPES measurements\cite{Miao14} detect $d_{xz}/d_{yz}$ orbital splitting  inside the superconducting phase in the parent LiFeAs as well as electron-doped LiFe$_{1-x}$Co$_x$As, with no magnetic phase nearby. Very recently, orbital ordering has also been observed\cite{Sonobe-unpub} inside the superconducting phase of the optimally doped and overdoped BaFe$_2$(As$_{1-x}$P$_x$)$_2$, far away from the magnetically ordered phase and in the same regime where the torque magnetometry detected $C_2$-symmetric spin response\cite{Kasahara12}.
These observations also raise the question of the interplay between orbital nematicity and superconductivity in the iron pnictides, which will be the subject of future study.

\begin{acknowledgments}
We are grateful to David S\'en\'echal for his help with the VCA method, 
and we are indebted to Andrei Chubukov for many fruitful discussions. 
We acknowledge discussions with Steve Kivelson, Yuji Matsuda, Abhay Pasupathy, 
Ian Fisher, Rafael Fernandes and Matthew Foster.
 This work was performed with partial support from the Welch foundation grant, 
 C-1818. A.H.N. acknowledges the hospitality of the Aspen Center for Physics 
 (NSF grant \#1066293) where part of this work has been performed.
\end{acknowledgments}

\appendix

\bibliography{\jobname}

\end{document}